# Open, Small, Rigmarole – Evaluating Llama 3.2 3B's Feedback for Programming Exercises

Imen Azaiz[1]*, Natalie Kiesler[2], Sven Strickroth[1]*, Anni Zhang[1]
[1] LMU Munich, Munich, Germany
[2] Nuremberg Tech, Nuremberg, Germany
`{imen.azaiz,sven.strickroth}@ifi.lmu.de`

**Abstract**—Large Language Models (LLMs) have been subject to extensive research in the past few years. This is particularly true for the potential of LLMs to generate formative programming feedback for novice learners at university. In contrast to Generative AI (GenAI) tools based on LLMs, such as GPT, smaller and open models have received much less attention. Yet, they offer several benefits, as educators can let them run on a virtual machine or personal computer. This can help circumvent some major concerns applicable to other GenAI tools and LLMs (e. g., data protection, lack of control over changes, privacy). Therefore, this study explores the feedback characteristics of the open, lightweight LLM Llama 3.2 (3B). In particular, we investigate the models' responses to authentic student solutions to introductory programming exercises written in Java. The generated output is qualitatively analyzed to help evaluate the feedback's quality, content, structure, and other features. The results provide a comprehensive overview of the feedback capabilities and serious shortcomings of this open, small LLM. We further discuss the findings in the context of previous research on LLMs and contribute to benchmarking recently available GenAI tools and their feedback for novice learners of programming. Thereby, this work has implications for educators, learners, and tool developers attempting to utilize all variants of LLMs (including open, and small models) to generate formative feedback and support learning.

## 1 Introduction

Generative AI (GenAI) and related tools have advanced rapidly. Beginning in 2022, computing education researchers investigated the underlying models and their capacities to solve programming problems, CS1 and CS2 exams [1]–[5]. Others were focusing on utilizing GenAI for instruction [6], [7] and supporting educators, for example by using explanations generated by Large Language Models (LLMs) [8], or to classify students' help requests [9]. Feedback is another huge potential of GenAI that has been explored with regard to its quality, characteristics, and feedback types [10]–[13]. GPT-4 Turbo, for example, has been shown to provide personalized feedback and perform significantly better than older versions [12]. Despite persisting weaknesses [12], this is particularly interesting for learners of programmers, who usually depend on feedback when they practice [14].

We are also seeing an increasing body of research on using GenAI in instructional settings, such as introductory programming courses and tutorials [15]–[18]. Survey studies with students revealed that computing students see the potential of GenAI tools, as they can help with understanding programming concepts, adapting and fixing code [19]–[21]. At the same time, students criticized the quality of the outputs and expressed concerns regarding ethical aspects [19]–[21]. Novice programmers particularly criticized the ease of use, availability, privacy issues, lack of integrity, and other aspects when chatting with ChatGPT-3.5 about their code [16]. These concerns are not entirely new [7], [22], [23], as GenAI models are perceived as a black-box. This lack of transparency for end-users can increase fear and should be avoided [24].

Open, and smaller models, e. g., Llama, may be an interesting alternative to counteract some of these challenges (e. g., data protection, privacy concerns, hidden changes, etc.). They can be run on a virtual machine or personal computer, and thus do not require expensive hardware. Both educators and students could use a lightweight model to generate formative feedback for programming exercises. However, there are only a few studies on the potential of open, small LLMs and their feedback for novice programmers [25], [26].

The **goal of this study** is to address this gap, and evaluate the feedback characteristics of an open, lightweight LLM, such as Llama 3.2 with 3 billion parameters. By analyzing Llama's feedback, we **contribute** to (1) understanding the feedback quality of open, small LLMs; (2) theory-building by refining feedback elements and categories; (3) benchmarking recent GenAI models; and (4) providing recommendations for educators and students interested in utilizing such models to generate feedback for programming tasks.

This paper is organized as follows: Section 2 reviews related work, and Section 3 details the methodology of the qualitative analysis. Section 4 presents the results, which are discussed in Section 5, followed by an examination of threats to validity in Section 6. Finally, the paper concludes with key findings and an outlook on future work.

## 2 Related Work

Providing feedback to novice learners who struggle to solve programming tasks is a well-known challenge in computing education. It is not surprising that automating feedback at scale has been subject to research and practice for decades (also using AI techniques) [27]–[29]. Before the broad availability of GenAI and related tools, most learning environments provided learners with the correct solution, a simple pass vs.

failed information, and knowledge about mistakes by pointing out compile errors and failed test cases [14]. Due to the advent of GenAI, and LLMs in particular, we have seen an increasing interest in utilizing these models for the generation of feedback [7], [10]–[13], [25], [26], [30]–[34].

For example, Balse, Valaboju, Singhal et al. [30] investigated the feedback generated by ChatGPT-3, and identified a high variability in the feedback's accuracy, correctness, and consistency. They suggest not to let students use the tool directly. Similarly, Hellas, Leinonen, Sarsa et al. [31] and Kiesler, Lohr, and Keuning [10] criticized the feedback of GPT-3.5, as it could not identify all issues of incorrect student code. Moreover, it would hallucinate issues [31], and provide misleading information for novices depending on the task [10]. Feedback on syntax errors seems to be more reliable and consistent across various LLMs [10], [31], [34].

With the more recent models (e. g., GPT-4 Turbo), we have seen several improvements regarding the quality of the feedback. Azaiz, Kiesler, and Strickroth [12] focused on the qualitative exploration of the model's feedback characteristics. They present a comprehensive set of categories referring to the feedback's content and structure, code representation, the correctness and correction type, suggested optimizations and coding style, and inconsistencies and redundancies [12]. In their replication of prior work on ChatGPT-3.5 [11], they also found that all feedback outputs were personalized – a novel characteristic.

As the feedback capabilities seem to be advancing in some ways, it is no surprise that most studies in the past two years focused on large LLMs and related proprietary tools, such as ChatGPT, Codex, and Copilot for programming education contexts. These tools, however, are associated with several challenges and limitations when it comes to their feedback [10], [30], [31], [34]. Privacy concerns, hidden changes, biases, intransparent training data, costs, and dependency add to the list of concerns [7], [15], [24]. Yet, small, open models and their feedback potential remain understudied.

To our knowledge, there is only a handful of studies evaluating the feedback generated by open LLMs [25], [26]. S Kumar, Adam Lones, Maarek et al. [25] examined the bug-fixing and feedback-generation abilities of CodeLlama and ChatGPT for Java programming assignments using JUnit tests. They noted a high variability in CodeLlama's output and some incorrect suggestions. Nonetheless, they conclude that LLMs "have the potential to address the extensively researched problem of generating effective formative feedback in CS1 education." ([25], p. 92).

Koutcheme, Dainese, Sarsa et al. [26] quantitatively explored the feedback of several LLMs. Among them were five powerful open-source models (CodeLlama with 7B, 13B, and 34B parameters; and Zephyr α and β with 7B parameters). They focused on the quality and relevance of code corrections via the categories "completeness", "perceptivity", and "selectivity" to determine how comprehensive or insightful the feedback is. The automated assessment of the feedback quality was conducted with GPT-4 as a judge, resulting in binary classifications. Hence, novel feedback characteristics remained unexplored. Yet, they recognize the potential of the smaller open-source models with fewer parameters, and the need for more research on how to utilize the smaller models in educational contexts [26].

The present work has the goal of addressing this gap by not only quantifying the performance of an open, small LLM but also qualitatively analyzing the feedback characteristics. This way, educators, tool developers, and students can gain valuable insights into the feedback qualities of models that can be run on a personal device.

## 3 Methodology

Our analysis is led by the following research question (RQ): *How can we characterize the feedback provided by an open, small LLM, such as Llama 3.2 (3B) if provided with a programming task description and a student solution as input?*

To answer this question, we requested and reused the dataset used in related work [11], [12]. The dataset comprises assignments and authentic student submissions from an introductory programming course in Java. Most of the students were in their first semester and majoring in computer science at LMU Munich a large university in Germany. The data was collected during the winter term 2021/22. About 900 students were registered for the course, 695 of them voluntarily consented that their data could be used for research. Students did not experience any advantages or disadvantages for (not) agreeing. The programming assignments were designed as voluntary homework. Students were asked to upload their solutions to the e-assessment system GATE [35], [36] to receive formative feedback. The course further included weekly exercise sessions conducted by student teaching assistants and voluntary peer (code) review among the students for one assignment per exercise sheet also administered using GATE [37], [38].

**Task Selection:** To support benchmarking various GenAI tools and models, we used the same two tasks and student submissions as related work [11], [12]. The tasks can be characterized as follows:

The first task was assigned in the second week of the course. It asked students to *"Write a Java application named SimpleWhileLoop that uses a WHILE loop to count and print all odd numbers from 1 to 10, and then print 'Boom!' (without quotation marks) afterward."*.

The second task was more advanced (week 7 of the course) and focused on object orientation, dynamic data structures, reference manipulation, and list traversing. It required students to: *"Implement the Queue interface according to the specification (in the interface) for a queue with the QueueImpl class by using a singly linked list."* Students should implement the Queue interface using an inner class for the nodes and the following five methods: `void append(int)`, `boolean isEmpty(int)`, `void remove()` (null operation on an empty queue), `int peek()` (returns the first value or the constant `EMPTY_VALUE` for an empty queue), and `int[] toArray()`. The specification for the methods was given as JavaDoc (in German) in the interface.

We reused the exact same sampled student submissions as in related work [11], [12]. Specifically, these comprise 33 pseudo-randomly sampled submissions for the *SimpleWhileLoop* task and 22 randomly sampled submissions for the *Queue* task. The sampled submissions represent approx. 10 % of the full dataset for each assignment.

**Feedback Generation:** To generate the feedback, Llama 3.2 was used with the following prompt (zero shot). The prompt is identical to the one used in related work [11], [12]:

```
[ASSIGNMENT INSTRUCTION]
Find all kinds of errors, including logical ones, and
provide hints for their correction or improvement,
including suggestions for code style.
[STUDENT SUBMISSION]
```

The feedback was generated three times for each submission with default settings in independent requests to address the probabilistic nature of LLMs and to investigate the consistency of the responses. 99 feedback texts for the *SimpleWhileLoop* and 66 for the *Queue* were generated.

**Feedback Analysis:** Before analyzing the qualitative feedback characteristics, all student submissions were assessed using the OpenJDK 11 compiler and unit tests. We also checked whether the student submissions were semantically correct by using human intelligence. The length of the feedback as well as the accuracy and recall of Llama 3.2's judgments were evaluated quantitatively.

Next, the feedback was analyzed using a qualitative thematic analysis method [39], [40]. The full student submission was used as a context unit. Each generated feedback text was treated as one coding unit, resulting in multiple codes being applied to one coding unit. The coding book developed in related work [12] served as a starting point for the inductive-deductive category-building process. The coding was conducted in the original language of the generated output. A significant amount of feedback for the *Queue* (64–69 %, i. e., 14 or 15 in each iteration) was generated in German – likely due to the German comments in the *Queue* interface's JavaDoc. If the feedback was not generated in English, it was translated into English after the analysis for this paper. Three researchers qualitatively analyzed the feedback. Unclear cases were discussed intensively until an agreement was reached. As a part of this process, we consulted related feedback taxonomies [28], [41], and studies [10]–[12].

## 4 Results

In this section, the feedback characteristics of Llama 3.2 are introduced. We begin by examining the overall structure of the outputs in terms of the content of the generated feedback, structure, language, clarity, and length.

Next, we present the qualitative findings of the deductive-inductive analysis of the feedback and the resulting categories (cf. Table 1). These refer to compliance with the task specification, the presentation of code, the quality of the corrections, types of suggestions, such as optimizations and code style, as well as inconsistencies, redundancies, and other characteristics.

Table 1. Coding book with definitions (examples are provided in the text where appropriate, ⋆: new, inductive category), cf. [11], [12]

| Category | Description |
| --- | --- |
| Compliance with task specification | |
| Compliance with spec. (CWAS) | Corrections or suggestions align with the provided instructions and task specification. |
| Code Representation | |
| Code snippet (CoSn) | Corrects small portions of the program suggesting a sequence of instructions. |
| Code with output (CWO) | Suggests improvements in the code with the corresponding output. |
| Inline code correction (ICC) | Feedback text contains student solution with inline comments (corrections and suggestions). |
| Correctness and Correction Types | |
| Partially correct correction/suggestion (PCCS) | Only some feedback components are correct, while other components introduce new issues (i.e., incorrect feedback or suggestions). |
| Only false correction/suggestion (OFCS) | Feedback contains only false corrections like non-existent errors or suggestions resulting in broken code. |
| ⋆ Only false error correction (OFEC) | Feedback contains only false corrections, such as non-existent errors or corrections resulting in broken code. Other suggestions (e. g., code style suggestions and general suggestions) are correct. |
| (Fault) localization (FL) | At least one bug is identified and localized, e. g., by citing code snippets, or describing them. |
| (Fault) localization correct (FLC) | All bugs are correctly identified and localized and are present in these locations. |
| Suggested Optimizations and Coding Style | |
| Optimization (OPT) | Suggests optimizations regarding the functionality of the program. |
| Code style suggestion (CSS) | Suggests improvements regarding readability, documentation, comments within the code, variable naming, etc. |
| Language suggestion (LCS) | Feedback contains translations and language related suggestions. |
| Inconsistencies and Redundancies | |
| Inconsistency (InC) | Recommendation does not correspond to the sample solution, or contradiction within the textual feedback. |
| Redundancy (RD) | Repeats the same suggestion in the same feedback or provides a suggestion that is already implemented in the code. |

## 4.1 Structure, Content, Language, and Length

### 4.1.1 Structure and Content

The output of Llama 3.2 exhibits almost a consistent structure across all feedback texts, although certain elements are present in some cases and absent in others. This seems to depend on the task. The feedback often starts with an introductory phrase like "Here's the corrected version of your code" (primarily for the *SimpleWhileLoop*) or "Here are the errors on the provided code". These introductory phrases are quite generic. We also found engaging sentences such as "That's a good start!" (translated) for the *Queue*.

The structure we present next varied depending on the task. Particularly for the *SimpleWhileLoop*, corrections are presented as a full code solution followed by (an enumerated) list of errors and suggestions containing code snippets themselves. This is vice versa for the *Queue* task. Moreover, enumerations contain textual descriptions of issues. These are often but not always accompanied by corrections as code snippets. In some instances, there are multiple enumerations categorized in sections, e. g., "Logical Errors" or "Improvements" or "code style". A full code solution often follows.

The feedback (for both tasks) sometimes ends with an abstract summary of the corrections. In a few cases, we also found such as "I hope this improved version helps." (translated). Overall, it does not seem like there is a consistent order of feedback elements.

All feedback outputs consistently contain text with code by mentioning keywords of the Java programming language, method/variable names, full code solutions, or code snippets. Moreover, all feedback is personalized to the student's submission reusing variable names or parts of the student code.

### 4.1.2 Language consistency and clarity

The feedback texts are not consistent in terms of the choice of language. A significant amount of feedback for the *Queue* was generated in German – likely due to the German comments in the *Queue* interface's JavaDoc. Many feedback texts are misleading and/or hard to understand, even for an experienced programmer/educator who knows what is wrong with the submission. The outputs also contain ambiguities, e. g., to indicate a possible resource leak:

> "*In the 'QueueImpl' class, when you're adding a new entry to an empty queue, you're setting both 'head' and 'tail'. However, in the 'remove()' method, you're only updating the 'head' reference. It would be more consistent to also update the 'tail' reference.*"

In other cases, incorrect terminology was used such as "class" instead of variable or reference. Not all texts are grammatically correct. Even words were misspelled (e. g., "methtodes"), and new words were invented (e. g., "Töpfname").

### 4.1.3 Length

The length of the generated feedback in terms of the number of words for both tasks is summarized in Table 2. The word count was determined by tokenizing the feedback text into words at whitespace ("\s+") and counting the resulting tokens. The feedback for the *Queue* is consistently longer across all metrics compared to the *SimpleWhileLoop*. Particularly, the overall median length for *Queue* (m = 573 words) is higher than that for *SimpleWhileLoop* (m = 364). This difference is statistically significant (Mann-Whitney UTest, U = 749.5, p < .00001, two-sided). The median is close to the arithmetic mean ($\bar{x}$) for both tasks ($\bar{x}$ = 374 for the *SimpleWhileLoop* and $\bar{x}$ = 586 for the *Queue*).

In an early test, an extremely long feedback text (8390 words) was generated for the *Queue*, with the same text repeating – this did not happen in the three runs.

**Table 2.** Length of the generated feedback (Llama 3.2) in terms of number of words (OA: overall for a task)

|  | SimpleWhileLoop | | | | Queue | | | | All |
| --- | --- | --- | --- | --- | --- | --- | --- | --- | --- |
|  | 1st | 2nd | 3rd | OA | 1st | 2nd | 3rd | OA |  |
| **Mean** | 382 | 365 | 377 | 374 | 615 | 603 | 541 | 586 | 459 |
| **Median** | 350 | 364 | 365 | 364 | 600 | 625 | 520 | 573 | 413 |
| **Min** | 227 | 239 | 273 | 227 | 366 | 352 | 364 | 352 | 277 |
| **Max** | 720 | 544 | 571 | 720 | 931 | 959 | 726 | 959 | 959 |

### 4.2 Compliance with Task Specification

In the following, we present the results of the qualitative analysis starting with the feedback's compliance with the task specification(s). All categories are summarized and defined in Table 1. The codes' frequencies are documented in Table 3.

Overall, 41 % of the feedback was categorized as compliant with the task requirements. The feedback for the *Queue* task complied with the task in 7 out of 66 cases (11 %); for the *SimpleWhileLoop* it is 60 instances out of 99 (61 %). These low numbers are closely related to incorrect corrections such as not considering odd numbers, not correcting a wrong capitalization of "BOOM", ignoring formatting issues, printing unnecessary text, appending at the front, or using exceptions when not allowed.

**Table 3.** Frequencies of all codes applied to both tasks *SimpleWhileLoop* and *Queue*

|  | SimpleWhileLoop (n=33) | | | | Queue (n=22) | | | | All (n=165) | |
| --- | --- | --- | --- | --- | --- | --- | --- | --- | --- | --- |
| **Char.** | **1st** | **2nd** | **3rd** | **OA** | **1st** | **2nd** | **3rd** | **OA** | **Sum** | **%** |
| *Content and Compliance with Task* | | | | | | | | | | |
| CWAS | 20 | 21 | 19 | 60 | 1 | 2 | 4 | 7 | 67 | 41 |
| *Code Representation* | | | | | | | | | | |
| FuCo | 33 | 33 | 33 | 99 | 15 | 17 | 15 | 47 | 146 | 88 |
| CoSn | 3 | 4 | 5 | 12 | 16 | 7 | 9 | 32 | 44 | 27 |
| CWO | 1 | 0 | 0 | 1 | 0 | 0 | 0 | 0 | 1 | 1 |
| ICC | 2 | 2 | 2 | 6 | 6 | 2 | 1 | 9 | 15 | 9 |
| *Correctness and Correction Types* | | | | | | | | | | |
| PCCS | 33 | 33 | 33 | 99 | 14 | 18 | 11 | 43 | 142 | 86 |
| OFCS | 0 | 0 | 0 | 0 | 8 | 4 | 11 | 23 | 23 | 14 |
| OFEC | 16 | 13 | 19 | 48 | 5 | 7 | 3 | 15 | 63 | 38 |
| FL | 33 | 30 | 30 | 93 | 22 | 22 | 22 | 66 | 159 | 96 |
| FLC | 7 | 4 | 2 | 13 | 0 | 1 | 1 | 2 | 15 | 9 |
| *Suggested Optimizations and Coding Style* | | | | | | | | | | |
| OPT | 15 | 17 | 18 | 50 | 12 | 13 | 13 | 38 | 88 | 53 |
| CSS | 32 | 33 | 33 | 98 | 13 | 16 | 13 | 42 | 140 | 85 |
| LCS | 4 | 2 | 3 | 9 | 0 | 0 | 0 | 0 | 9 | 5 |
| *Inconsistencies and Redundancies* | | | | | | | | | | |
| InC | 31 | 30 | 28 | 89 | 16 | 18 | 12 | 46 | 135 | 82 |
| RD | 21 | 17 | 22 | 60 | 15 | 12 | 13 | 40 | 100 | 61 |

### 4.3 Code Representation

Almost all generated feedback texts contain code in the form of full code solutions (FuCO) and/or code snippets (CoSn), as summarized in Table 3. Every feedback for the *SimpleWhileLoop* contains a full code solution (100 %). For the *Queue*, this applies to 84 % of the feedback. Overall, code snippets were found in 27 % of the feedback. They are more prevalent in the *Queue* task, appearing in 32 out of 66 instances, compared to 12 out of 99 cases in the *SimpleWhileLoop* task. Only one feedback for the *Queue* task neither contained a full code solution nor code snippets.

Code with its corresponding output (CWO) was identified in only one feedback for the *SimpleWhileLoop*. Student code with inline comments (ICC) occurred 6 times for the *SimpleWhileLoop* (9 %), and 9 times for the *Queue* (14 %). We did not find any code snippets with comments prompting the student to fill a gap based on the LLMs' instructions (defined as "CoSnI" in related work [12]).

### 4.4 Feedback Correctness and Corrections

An important aspect of the analysis of the feedback's correctness was the classification performance of Llama 3.2, i. e. assessing correct submissions as correct and incorrect ones as incorrect. We evaluated this performance by using different metrics (cf. Table 4). To calculate these metrics, we analyzed the correctness of the students' submissions with human intelligence. As a result, we found that the majority (57 %) of submissions in response to the *SimpleWhileLoop* are fully correct and 90 % are syntactically correct. However, only three submissions for the *Queue* are completely correct and 64 % are syntactically correct.

Table 4. Overview of Lama 3.2's classification performance

| | SimpleWhileLoop | | | | Queue | | | | All |
|---|---|---|---|---|---|---|---|---|---|
| **Metric** | **1st** | **2nd** | **3rd** | **OA** | **1st** | **2nd** | **3rd** | **OA** | |
| **Accuracy** | .52 | .52 | .52 | .52 | .86 | .91 | .86 | .88 | .66 |
| **Precision** | .60 | .50 | .20 | .44 | – | 1.00 | – | 1.00 | .48 |
| **Recall** | .18 | .25 | .07 | .17 | .00 | .33 | .00 | .11 | .16 |
| **Specificity** | .88 | .76 | .80 | .81 | 1.00 | 1.00 | 1.00 | 1.00 | .91 |

The overall accuracy of Llama's classification (ratio of correct results to all results) is .66. The accuracy is higher for the *Queue* (.88) compared to *SimpleWhileLoop* (.52).

The precision (ratio of true positive results to all positive results) for the *Queue* could only be calculated for the second iteration due to the absence of false positives in other iterations. The resulting value is 1.0. In contrast, the precision for the *SimpleWhileLoop* ranges from .2 to .6. The recall (ratio of true positive to all positive results) is generally low for both tasks ranging from .0 to a maximum of .33 for the *Queue* task. This result indicates challenges in recognizing correct submissions as correct. The specificity (ratio of true negative to all negative results) is very high across both tasks (.91), and perfect (1.0) for the *Queue* task. Specificity ranges between .76 and .88 for the *SimpleWhileLoop*.

### 4.4.1 Correctness and Correction Types

In general, Llama 3.2's feedback always includes corrections and suggestions for improvements. However, no feedback is completely correct or fixes all issues of the student submission. Every feedback in response to the *SimpleWhileLoop* and the majority (65 %) for the *Queue* is only partially correct (PCCS, overall 86 %, cf. Table 3). Notably, one-third of the feedback for the *Queue* is completely incorrect (OFCS); as there are no cases for the *SimpleWhileLoop*. A newly identified, reoccurring characteristic involves feedback with completely all error corrections being incorrect, but all other code style and general suggestions are correct, or they do not introduce new issues (OFEC). This code applies to 38 % of the feedback texts overall and occurs more frequently in the *SimpleWhileLoop* (48 % vs. 23 %).

Llama 3.2 localized bugs (FL) using method/variable names or short quotes of code in 96 % of all cases (94 % *SimpleWhileLoop* and 100 % *Queue*). Usually, the right position of an issue is identified, but errors are incorrectly attributed. Overall, 9 % of the fault localization attempts are fully correct (FLC), with just two cases in the *Queue* task and 13 cases in the *SimpleWhileLoop* task. For the *Queue*, there are cases where the error identification, problem description, or textual correction is incorrect, but the suggested correction in a code snippet is correct.

Common student errors in the SimpleWhileLoop relate to a wrong capitalization of the word "Boom!", the loop, or printing even numbers. These errors were often correctly localized. However, "errors" such as the initialization of the counter variable (0 vs. 1), wrong loop condition (`x < 10` vs. `x <= 10`), missing new lines, or non-existing syntax errors like a missing semicolon directly after `if` or `while` were hallucinated and subsequently "corrected" by the model.

For the *Queue*, Llama 3.2 seems to be able to detect many different kinds of issues, e. g., additional output text, resource leaks, missing `null` checks in the `remove` method, use of Java's `LinkedList` or `ArrayList`, and off-by-one errors in loop conditions, however, not consistently. Yet, there are many incorrect corrections such as throwing an exception in the constructor if the length is zero, using Java's `ArrayList` or an `int` array, or hallucinated syntax errors. Syntactic errors such as `if (size = 0)` are hardly ever detected. Notably, all but one full code corrections (FuCo) for the *Queue* are syntactically incorrect. For an almost empty *Queue* submission, where the student stated that the task was too difficult, Llama 3.2 did not take these comments up and did a "normal" correction.

Overall, many corrections are literally nonsense, for example:

- "Use Markdown formatting for code blocks"
- "The class should be written in font size 12" (translated)
- "The test for the empty list should use a 'while' loop to ensure that all elements in the list are found." (translated, in `isEmpty()`)
- "if (size <= 1) { //just making sure size 0 works" (in `isEmpty()`)
- "The class 'Queue' uses obsolete Java property names ('public', 'private', 'protected', 'final')." (translated)
- `if (!this.isEmpty() && this.head == null) {`
- `if (this == null) {` (in `isEmpty()`)
- "Removed unnecessary semicolon: After 'System.out.println(i);', there was a semicolon in the original code, which is not needed in this case."

### 4.5 Suggested Optimizations and Coding Style

About half of all feedback texts (53 %) generated by Llama 3.2 include suggestions for optimization (OPT, cf. Table 2). This is consistent for both tasks and across all three iterations. For the *Queue*, there are many reasonable suggestions, e. g., considering thread-safety, adding `@Override` (often added without being explicitly mentioned) or introducing a `tail` field for $\mathcal{O}(1)$ addition. For both tasks, avoiding magic numbers, and implementing the `toString()` method for debugging are suggested. Other recommendations were unreasonable, such as using exceptions (*Queue* and *SimpleWhileLoop*, but not allowed), introducing backward references for the *Queue*, adding Java generics in the interface, checking that a primitive `int` parameter is not `null`, or removing the "unnecessary" `toArray()` method. The `EMPTY_VALUE` constant was also quite often replicated in the *QueueImpl* class.

Code Style Suggestions (CSS) are included in 85 % of all generated feedback texts. Specifically for the *SimpleWhileLoop* task, all but one case contains such suggestions. Only about half of the feedback for the *Queue* task contains CSS elements. The suggestions focus on using comments, descriptive names, and improving code readability. Notably, the corrected full code (FuCo) for the *SimpleWhileLoop* often includes JavaDoc. For both tasks, many corrections include inline comments highlighting the changes. Hence, these only serve the correction but not the documentation requested in the feedback. Also, useless comments such as `return head == null; // Improved description of the condition` are suggested.

Language related suggestions such as using English variables names are very rare and only found for the *SimpleWhileLoop* (9 %).

### 4.6 Inconsistencies and Redundancies

The feedback generated by Llama 3.2 contains inconsistencies in 82 % of the cases (cf. Table 3). Inconsistencies manifest, among others, in mislabeled section headings. Often code style suggestions are included within the "errors" subdivision. Other inconsistencies were observed within code snippets or explanations, for example:

*"When removing an element from the queue, the head should be updated to the next node in the list (after the one being removed). [...] The current implementation incorrectly updates the head to the first node after the head."*

*"It would be more consistent to use camelCase for variable names (e.g., `size` instead of `size`)."*

We identified many redundancies in 61 % of the cases. Among them were many code style recommendations to use consistent indentation or to camelCase names for perfectly indented submissions with camelCase variable and method names. In other cases, nearly identical advice is repeated several times.

## 5  Discussion

GPT-3.5 and GPT-4 are generic LLMs that have been shown to perform quite well in providing feedback to introductory programming tasks and student solutions [11], [12]. One of the benefits of small LLMs (e. g., Llama 3.2 with 3B parameters) is that they can be used without a high-end GPU. For example, on a recent 13th Gen Intel i5-13500 CPU (20 core, 2.5 GHz), one feedback message can be generated in about 2 minutes. Hence, educators can generate feedback for an assignment overnight. Even students can use it on their devices. These are reasonable use cases (cf. [42], [43]). Larger LLMs may cause a significant slowdown without expensive GPUs. So, we focused on a small LLM, i. e., the 3B model, in alignment with the anticipated uses.

The analysis of the feedback characteristics started with the reuse of the coding book from related work [12]. Overall, the categories applied well to Llama's feedback. However, we found no instances of the following categories: *only correct corrections/suggestions, completely correct correction,* and *code snippet with instruction* [12]. At the same time, we built a new, inductive category: *Only false error correction* (cf. Table 2). Additional categories, such as *nonsense* or *language error* may be advisable in future work.

The structure of the feedback is comparable to that of GPT-4 [12]. GPT-4 mostly provided a summary of the task in the beginning, and subsections were more clear and consistent. The median length of Llama's feedback is about 16 % longer for the *SimpleWhileLoop* and about 22 % longer for the *Queue*.

Regarding the correctness of the feedback, Llama 3.2 showed a perfect specificity (1.0) in the context of the *Queue* task. It should be noted though that 19 of the 22 student submissions for the *Queue* are incorrect. Classifying those as incorrect thus has a chance of 86 % of being correct. Thus, the very low recall (.11) must be considered.

This is somewhat different for the *SimpleWhileLoop* task, where the incorrect submissions are not dominant. Specificity remains high (.81), but recall is also low (.17). These numbers indicate a frequent misjudgment of correct submissions. This raises the question of whether the high specificity genuinely reflects classification performance. It may also reflect a bias in the dataset. However, the submissions were selected (pseudo-)randomly, and a real scenario is also unlikely to have a balanced dataset.

Output formatting was reported as an issue for older models (e. g. Codex [1], [31] and GPT-3.5 [11], [31]), but could not be confirmed with GPT-4 Turbo [12], is an issue for this recent and small model. An analysis of the feedback of other open, small LLMs in related work [26] has shown that the smallest model (CodeLlama-7B) delivers 18 % comprehensive feedback for programming tasks (i. e., feedback that identifies all issues, without providing nontruthful information). Comparing these results to our findings reveals the deficits of Llama 3.2, as we could not find any example of a completely correct correction (CCC). Regardless, all of the open, small models perform much worse than GPT-3.5 Turbo, GPT-4 [26], and GPT-4 Turbo [12].

It is also interesting to compare the performance of Llama 3.2's with (first semester) student peer (code) review, which is also often used to address the scalability issue of providing personalized feedback, as there seem to be similarities and differences in their quality: Accuracy (Llama 3.2: .66; peer review overall .64 and ranging from .46 to .71 in [37]) and specificity (Llama 3.2: .91 vs. peer review .85 in [37]) seem to be comparable, but the recall seems to be significantly higher in peer review (Llama 3.2: .16 vs. .42 in [37]). Also, students often fail to identify subtle errors such as formatting issues and typos in class or method names [38]. Furthermore, students often only provide very short reviews (e. g., median length of 13 words in [37] and 14 words in [44]), but only provided incorrect corrections in about 15 % of the cases [38].

Finally, it is worth discussing the structure of Llama 3.2's feedback in the context of novice programmers as potential users. Almost all feedback texts for the *SimpleWhileLoop* start with a full code solution. This is certainly not ideal for feedback intended to guide students toward the solution. Also, not reacting on student comments asking for help, is nothing a human tutor would do. Still, we found motivational statements, but very few. Moreover, feedback with 81 % inconsistencies, 61 % redundancies, and many incorrect corrections (FLC is 9 %) creates a high cognitive load for any reader, particularly novices. Grammar and spelling issues along with nonsense text exacerbate this issue. Therefore, using Llama 3.2 as an educator to prepare formative feedback for students does not seem advisable. Similarly, computing students with limited programming experience should not consult this model for feedback.

## 6   Threats to Validity

LLMs such as Llama are probabilistic models that predict new tokens based on the context and already emitted tokens. Hence, LLMs can produce different outputs for identical settings and input. To counteract this limitation, we generated the feedback three times so we do not base our analysis on outliers. However, we cannot predict how a fourth generation would look like.

Another limitation is due to the language(s) used in the prompt, the programming language, and the selected assignments. LLMs are known to produce different outputs depending on the requested language. Finally, it should be noted that despite Llama being labeled as "open", we do not know what training data has been used.

## 7 Conclusions and Outlook

This work presents a qualitative analysis of feedback generated by Llama 3.2 in response to a programming task description and a student solution. We reused the dataset, methodology, and category system developed in related work [12] to explore Llama 3.2's feedback characteristics. Reusing data yields comparable results, and contributes to benchmarking various LLMs.

Accordingly, our results show that the deductive-inductive category system developed in related work can be applied to Llama 3.2 with only a few adaptations. Overall, the feedback quality is limited, compared to other LLMs, as many errors were not identified (e. g., syntax errors), many corrections were wrong (86 % is only partially correct), there are no completely correct corrections, about 82 % of the feedback contains inconsistencies, and about 38 % of the error correcting feedback resp. 14 % overall was complete nonsense. Nonetheless, Llama 3.2 can detect the same types of errors and suggest similar helpful optimizations as large LLMs such as GPT-4 [12]. To conclude, Llama 3.2's feedback performance proved to be impressive for a small model, but it was devastating at the same time. At this point, a small model, such as Llama 3.2 does not seem suitable for introductory programming contexts – neither for students seeking nor educators trying to get help in generating it.

These insights are crucial for educators, tool developers, and students considering the use of Llama 3.2. Despite the potential benefits (e. g., data protection, version control, etc.), it does not (yet) seem recommendable to use the open, small LLMs. Nonetheless, future work is needed to investigate other smaller models, specialized in code, such as CodeLlama (7B) or qwen2.5-coder (.5B), as they also seem to have potential. Such studies must explore not only quantitative performance metrics but also qualitative elements to better understand the strengths and weaknesses of GenAI tools. Ideally, future work reuses datasets from prior work so we can continue benchmarking relevant tools, and get the much-needed insights into their quality and limitations. This is particularly important due to the rapidly advancing technologies, and the related development of educational tools based on GenAI.

## 8 Acknowledgements

The authors thank the students who allowed us to use their submissions for this research. This research was supported by the German Federal Ministry of Education and Research (BMBF), grant number [16DHBKI013].

## 10   Authors

**Imen Azaiz** is a research assistant at the Institute for Informatics of Ludwig-Maximilians-Universität München, Germany (Oettingenstraße 67, 80538 München, Germany). She received her Master's degree in Computer Science from LMU Munich in 2020 and is a certified Artificial Intelligence Engineer. She works in the AIM@LMU project and is pursuing a Ph.D. in Computer Science. Her research interests are Technology-Enhanced Learning and Learning Analytics. (email: imen.azaiz@ifi.lmu.de, ORCID: https://orcid.org/0009-0005-6458-4169).

**Natalie Kiesler** is a professor of teaching and learning in higher education at Nuremberg Tech's Faculty of Computer Science (Hohfederstraße 40, 90489 Nuremberg). Previously, she was a senior researcher at DIPF and a lecturer at Goethe University Frankfurt, where she earned her doctorate in Computer Science in 2022. Her research focuses on programming competency, learning environments, and feedback in university-level education. Current projects include generative AI, feedback, open science, and equity in education. She has received several awards, including the Hessian University Award for Excellence in Teaching. Natalie serves in various leadership roles in academic conferences and is an active reviewer for several journals and conferences. (email: natalie.kiesler@th-nuernberg.de, ORCID: https://orcid.org/0000-0002-6843-2729).

**Sven Strickroth** is a professor of Technology-Enhanced Learning at the Institute for Informatics of Ludwig-Maximilians-Universität München, Germany (Oettingenstraße 67, 80538 München, Germany). He graduated in Computer Science at Clausthal University of Technology and received his doctorate in Computer Science in 2016 at Humboldt-Universität zu Berlin, Germany. His research interests include E-Assessment, Learning Analytics, and Computer-Supported Collaborative Learning, primarily but not limited to the context of computer science education. He has received several awards, including the Award for Excellence in Teaching from the Bavarian Ministry of Research and Arts. He is a co-founder of the German workshop series on automated assessment of programming assignments and is a board member of the special interest group Educational Technologies of the German Computer Society (GI). (email: sven.strickroth@ifi.lmu.de, ORCID: https://orcid.org/0000-0002-9647-300X).

**Anni Zhang** is a student at the Institute for Informatics of Ludwig-Maximilians-Universität München, Germany (Oettingenstraße 67, 80538 München, Germany). She is about to complete her Bachelor of Science in Computer Science. (email: anni.zhang@campus.lmu.de, ORCID: https://orcid.org/0009-0005-6506-3779).